\documentclass{emulateapj}  
\slugcomment{ApJ Letters, accepted}

\newcommand{\cms}{$~{\rm cm~s^{\scriptscriptstyle -1}}$}
\newcommand{\smy}{M$_{\odot}{\rm yr^{\scriptscriptstyle -1}}$}

\newcommand{\ergs}{$~{\rm erg~s^{\scriptscriptstyle -1}}$}

\newcommand{\kmsMpc}{${\rm km~s^{\scriptscriptstyle -1}~Mpc^{\scriptscriptstyle -1}}$}
 
\newcommand{\grb}{GRB~060218}
\newcommand{\sn}{SN~2006aj}
\newcommand{\snbw}{GRB~980425/SN~1998bw}
\newcommand{\sndh}{GRB~030329/SN~2003dh}
\newcommand{\snaj}{GRB~060218/SN~2006aj}
\newcommand{\av}{$A_V = 0.46~{\rm mag}$}

\newcommand{\mvhost}{$M_{V,host} = -16.0~{\rm mag}$}

\newcommand{\othree}{[\ion{O}{3}]~$\lambda5007$}
\newcommand{\otwo}{[\ion{O}{2}]~$\lambda3727$}
\newcommand{\ntwo}{[\ion{N}{2}]~$\lambda6584$}
\newcommand{\sitwo}{\ion{Si}{2}~$\lambda6355$}

\begin{document}

\title{Early-Time Photometry and Spectroscopy of the Fast Evolving
 SN~2006aj Associated with GRB~060218\footnote{Observations reported
 here were obtained at the MMT Observatory, a joint facility of the
 Smithsonian Institution and the University of Arizona.}}

\author{M.~Modjaz\altaffilmark{1}, 
K.~Z.~Stanek\altaffilmark{2}, 
P.~M.~Garnavich\altaffilmark{3},
P.~Berlind\altaffilmark{4},
S.~Blondin\altaffilmark{1},
W.~Brown\altaffilmark{1},
M.~Calkins\altaffilmark{4},
P.~Challis\altaffilmark{1},
A.~M.~Diamond-Stanic\altaffilmark{5},
H.~Hao\altaffilmark{1},
M.~Hicken\altaffilmark{1}, 
R~P.~Kirshner\altaffilmark{1},
J.~L.~Prieto\altaffilmark{2}
}
 \altaffiltext{1}{Harvard-Smithsonian Center for Astrophysics,
 60 Garden Street, Cambridge, MA, 02138; mmodjaz,kirshner,mhicken,pchallis,sblondin,wbrown,hhao@cfa.harvard.edu.}
\altaffiltext{2}{Dept. of Astronomy, The Ohio State University, Columbus,        OH 43210;kstanek,prieto@astronomy.ohio-state.edu}

\altaffiltext{3}{Department of Physics, 225 Nieuwland Science Hall, University of Notre Dame, Notre Dame, IN 46556; pgarnavi@nd.edu}
\altaffiltext{4}{Smithsonian Institution, Fred Lawrence Whipple Observatory, 670 Mount Hopkins Road, P. O. Box 97, Amado, AZ 85645}
\altaffiltext{5}{Steward Observatory, University of Arizona, 933 North Cherry                       
Avenue, Tucson, AZ 85721 }

\begin{abstract}

We present early photometric and spectroscopic data on the afterglow
of GRB~060218 and report the evolution of the underlying supernova
2006aj. Our data span a time-range of 4 days to 23 days after the GRB
and clearly establish that SN~2006aj is a fast-evolving broad-lined
Type Ic SN with an extremely short rise-time ($\sim$10 days) and a
large optical peak luminosity ($M_V = -18.7$ mag).  The supernova
properties are deduced well since the GRB afterglow does not
contribute a significant amount to the total light output. The spectra
show broad lines indicative of large expansion velocities, but are
better matched by those of SN~2002ap and SN~1997ef (not associated
with a GRB) than those of the proto-typical GRB-related SN~1998bw. We
refine the redshift estimate to $z=0.03351~\pm~0.00007$. The
host-galaxy is a low-metallicity dwarf galaxy (with \mvhost), similar
to host-galaxies of other GRB-associated SNe.
\end{abstract}


\section{INTRODUCTION}\label{intro_sec}

The connection between long-duration gamma-ray bursts (GRBs) and
supernovae (SNe) that arise from the core collapse of very massive
stars stands on firm footing (see, e.g., \citealt{stanek05rev} for a
recent review). Circumstantial evidence includes the location of GRBs
near sites of massive star formation \citep{bloom02} and the detection
of unusual ``bumps'' in the GRB afterglow light curves that mimic a SN
light curve and its colors (e.g., \citealt{bloom99,garnavich03}). More
importantly, there are two cases of direct associations: the temporal
and spatial coincidence between SN~1998bw and GRB~980425
\citep{galama98} and the metamorphosis of the GRB~030329 spectrum into
that of a supernova, called SN~2003dh
\citep{stanek03,hjorth03,matheson03,kawabata03,kosugi04}. Based on
spectra, SN~1998bw and SN~2003dh were classified as peculiar Type Ic
SNe (SNe Ic), that is, core-collapse SNe that showed no hydrogen nor
helium in their spectra, but unusually high expansion velocities
requiring high explosion energies
\citep{galama98,iwamoto98,patat01,mazalli03}, in line with the
collapsar model for long-duration GRBs
\citep{woosley93,macfadyen99}. There is some evidence of
spectroscopically confirmed supernovae for other GRBs, however with
less confidence \citep{dellavalle03,malesani04}. Spectral confirmation
of GRB-related SNe is crucial since spectra provide understanding of
the velocity and nucleosynthesis structure of the ejecta and give an
insight into the energetics of the SN explosion. Here we report on the
spectral and photometric evolution of \snaj.
 
On 2006 February 18, at 03:34:30 UT, the Burst Alert Telescope (BAT)
onboard the $Swift$ Gamma-Ray Burst Explorer detected the bright
GRB~060218 \citep{grbtrigger}. $Swift$ XRT and $Swift$ UVOT also
detected its afterglow in the X-Ray \citep{grbtrigger,grbxrt} and
optical bands \citep{grbtrigger}, that lead to a precise determination
of the optical counterpart's position as $\rm{RA}=03^h 21^m 39^s.71$
and $\rm{DEC}=+16\arcdeg 52\arcmin 02{\farcs}6$ (equinox J2000.0)
\citep{grbuvot}. \grb~lasted about 2000 seconds \citep{grblast},
establishing it as one of the longest GRBs, and had peculiar gamma-ray
and X-ray afterglow properties \citep{grbgehrels}. Due to the unusual
properties and relative brightness of \grb, extensive and rapid
follow-up observations in all wavelength bands ensued around the
globe. Spectroscopic observations of the host galaxy, that had been
detected in pre-burst images by SDSS \citep{cool06}, and of the
optical transient (OT) were undertaken by several groups. Their
spectra revealed a blue continuum due to the afterglow light, narrow
host galaxy emission lines at a redshift of $z$=0.033 \citep{grbz} and
broad spectral features characteristic of a supernova
(\citealt{masetti06}, \citealt{soderberg06gcn}, \citealt{mazzali06},
\citealt{fugazza06}, \citealt{fatkhullin06}), which was designated
SN~2006aj (\citealt{soderberg06}, \citealt{mirabal06},
\citealt{fugazza06}). SN~2006aj was visible at $\Delta T \sim 2$ days
(where $\Delta T$ is time after burst) in the restframe of the GRB
\citep{masetti06,mirabal06}, much earlier than in \sndh~(where the SN
was visible at $\Delta T \sim$ 8 days in the rest frame). With a
redshift of $z$=0.033, this is the lowest-redshift GRB/SN after
\snbw~($z=0.0085$), and closer than \sndh~at $z=0.1685$.

Due to its proximity and early emergence of the SN, a densely
time-sampled and high-quality study of \snaj~can shed light on
GRB-related core-collapse SNe, their explosion mechanism and
progenitor history. Here we present such a dataset of spectroscopic
and photometric observations.


\section{OBSERVATIONS}\label{obs_sec}

After the announcement of the SN-GRB association via the GCN, we
promptly established our monitoring program of \snaj. Spectra and
photometry were obtained on a nearly nightly basis starting UT 2006
February 22 with the 6.5m Multiple Mirror Telescope (MMT) and the 1.5m
Tillinghast and 1.2m telescopes at the Fred Lawrence Whipple
Observatory (FLWO). The spectrographs utilized were the Blue Channel
\citep{schmidt89} at the MMT, and FAST \citep{fabricant98} at the FLWO
1.5m telescope. All optical spectra were reduced and calibrated
employing standard techniques in IRAF\footnote{IRAF is distributed by
the National Optical Astronomy Observatory, which is operated by the
Association of Universities for Research in Astronomy, Inc., under
cooperative agreement with the National Science Foundation.} and our
own IDL routines for flux calibration.

All of our photometry data were obtained with the FLWO 1.2-m
telescope. In this paper we report on 26 $V$-band points obtained
between 2006 February 22 and March 13 UT, i.e. between 4 and 23 days
after the burst. The light curve was extracted using the ISIS2 image
subtraction package \citep{alard00}. To obtain absolute calibration,
we observed Landolt standards \citep{landolt92} on 2006 March 4
UT. The derived transformation coefficients and color terms were used
to calibrate a sequence of 9 stars near \sn. For future references and
cross-calibrations, this transformation gives $V= 15.21 \pm 0.01$ mag
for the SDSS star at $\rm{RA}=03^h 21^m 42^s.77$ and
$\rm{DEC}=+16\arcdeg 51\arcmin 39{\farcs}46$ (equinox J2000.0). The
absolute photometric calibration is thought to be accurate to
$\sim$5\%. We note that our calibration yields comparison star
magnitudes in the SDSS $griz$ system (using the transformation
equations of \citealt{lupton05}) that are fainter than given in
\citet{cool06}. The offsets in the respective bands are: +0.81 mag
($u'$), +0.40 mag ($B$), +0.27 mag ($V$), +0.20 mag ($R$), +0.23 mag
($r'$) and +0.15 mag ($i'$). Using the same transformation, we obtain
$V=20.21$~mag for the host galaxy, using data from \citet{cool06} and
our own Landolt calibrations.

\section{RESULTS}\label{res_sec}

\subsection{\snaj}

\begin{figure}
\epsscale{.80}
\plotone{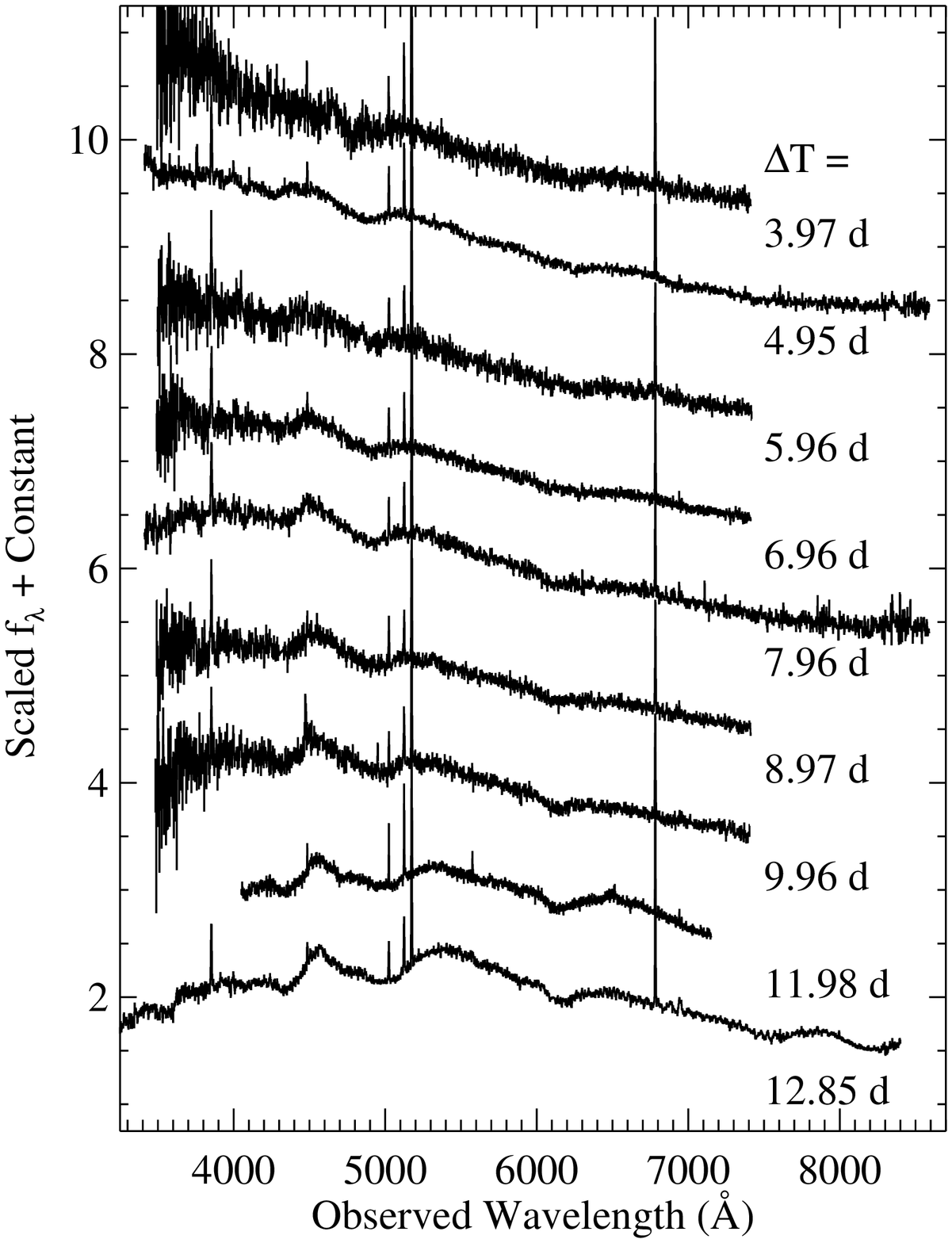}
\caption{Observed spectral evolution of the GRB~060218/SN~2006aj, from
February 22.12 UT (3.97 days after the burst) to March 3.00 UT (12.85
days after the burst), dereddened by \av~of Galactic extinction
\citep{schlegel98}. The spectra consist of a powerlaw continuum,
typical of GRB afterglows, and the broad features characteristic of a
peculiar, broad-lined Type Ic supernova. The narrow emission lines
originate from the host galaxy at a redshift of $z$=0.0335. }
\label{mont_fig}
\end{figure}

Figure~\ref{mont_fig} shows a montage of our spectra, corrected for a
Galactic extinction of \av~\citep{schlegel98}. We do not correct for
host-galaxy extinction, which is constrained to be small, namely
$E(B-V)\sim$0.04 mag from the equivalent widths of Na~I~D lines in
high-resolution spectra \citep{guenther06}.  On top of a smooth
powerlaw continuum that is typical of GRB afterglows are undulations
characteristic of a ``broad-lined'' Type Ic supernova: broad blends of
\ion{Fe}{2} and \sitwo, while no hydrogen or helium absorption lines
are detected. The SN signatures are visible in our earliest spectrum
at observed $\Delta T$=3.97 days and become stronger as a function of
time. The narrow lines are common nebular lines and identified as
H$\alpha$, \othree, H$\beta$, and \otwo~at $z=0.03351~\pm~0.00007$
arising from \ion{H}{2} regions in the host galaxy. Assuming
$\Omega_M$ = 0.3, $\Omega_{\Lambda}$ = 0.7 and $H_o$ = 72~\kmsMpc,
this corresponds to a luminosity distance of 143 Mpc. Using the value
for the fluence detected by $SWIFT$ \citep{campana06}, GRB~060218
appears to be an underluminous event in gamma rays: the isotropic
equivalent energy amounts to $E_{iso} = 6 \times 10^{49}$ erg
(extrapolated to 1$-$10,000 keV band in the GRB rest frame), which is
$\sim~10^{-2} - 10^{-1} $ that of cosmological GRBs (e.g.,
\citealt{ghirlanda05,friedman05}), but larger than for \snbw. The
gamma-ray peak luminosity is $L_{peak} =5 \times 10^{46}$ \ergs~(see
also \citealt{grbfluence,campana06}).

\begin{figure}
\epsscale{1.0}
\plotone{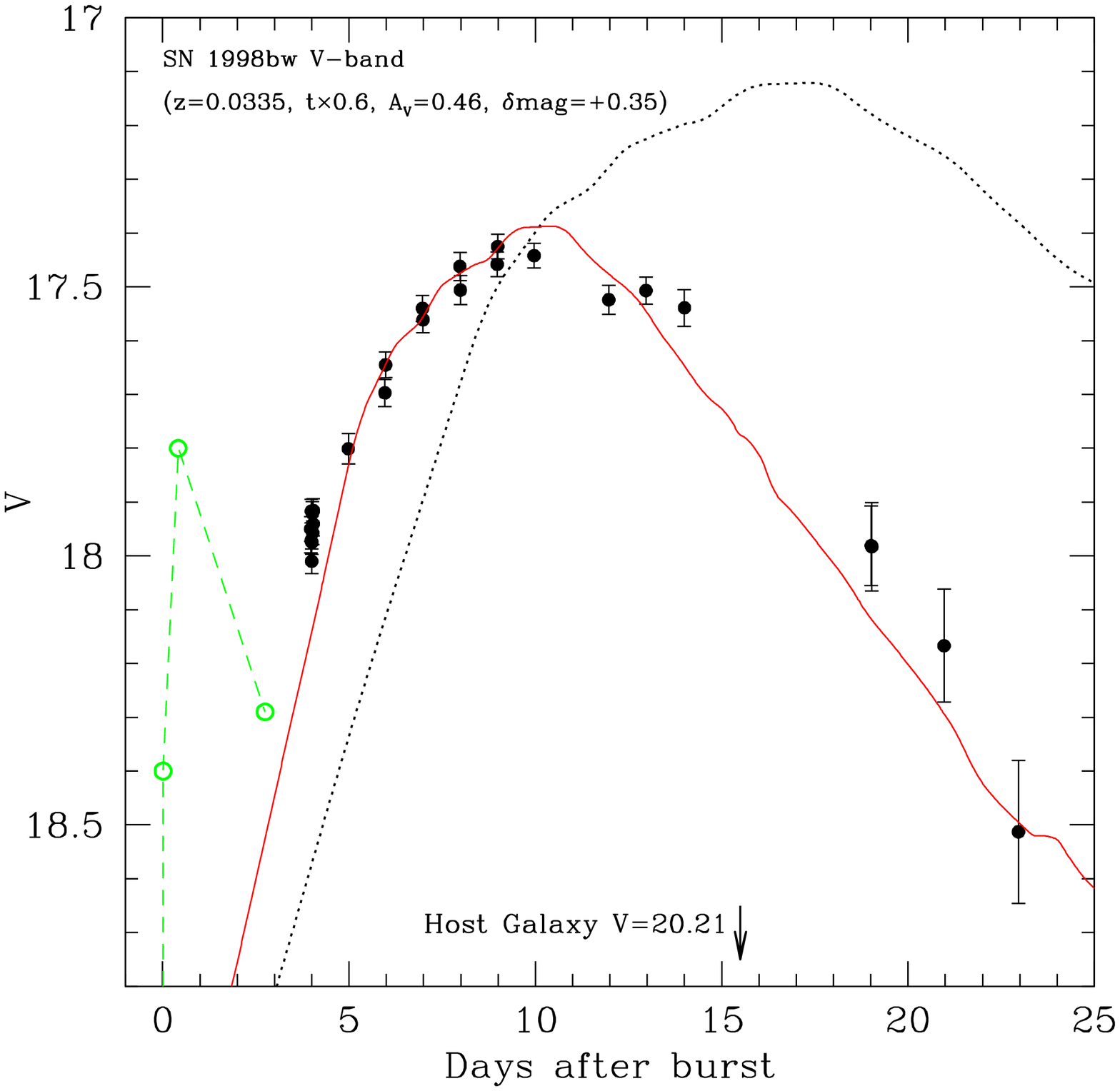}
\caption{Observed $V$ light curve of \snaj~based on the FLWO 1.2m data
(filled points) and $Swift$ UVOT data taken from the GCN (open
points). Superimposed is the $V$ light curve of SN~1998bw, k-corrected
and time-dilated to $z$=0.0335, stretched by a factor of 0.6, dimmed
by 0.46 mag of Galactic extinction, the SN host galaxy contribution
added, and shifted by 0.35 mag to match that of SN~2006aj (solid
line). No correction for host galaxy extinction has been applied to
SN~2006aj or the comparison light curve of SN~1998bw. See
\S~\ref{res_sec} for details. [\it{See the electronic edition of the
Journal for a color version of this figure.}\rm] }
\label{vlc_fig}
\end{figure}

Figure~\ref{vlc_fig} presents our early $V$-band light curve for
\snaj~(filled circles). In addition to our FLWO 1.2-m data, we have
added some data from the literature to extend our time coverage,
namely four early $V$-band points from $Swift$-UVOT as reported by
\citet{grbuvot,grbuvot2,grbxrayfull}.  \citet{campana06} present
$SWIFT$ early-time data ($\Delta T < 10^6$ sec or 11.6 days) of
\snaj~and argue for tantalizing evidence of an observed thermal
shock-breakout along with the GRB afterglow thousands of seconds after
the collapse of the core. Our $V$-band data clearly show a second peak
and the shape of a "supernova bump". The combined GRB afterglow/shock
breakout decline very quickly and the SN component dominates the OT
light early on ($\Delta T \geq 3$ days) and certainly when our spectra
were obtained. At those times, \sn~is $6-13$ times brighter than the
host galaxy. To compare the behavior of this bump, we used the
$V$-band light curve of SN~1998bw \citep{galama98}, shifted to
$z=0.0335$ and dimmed by $A_V=0.46\;$mag due to the Galactic
extinction \citep{schlegel98} (dotted line in
Figure~\ref{vlc_fig}). We find that the SN~2006aj associated with
GRB\,060218 evolved much faster than SN~1998bw. Indeed, we find a good
fit to our $V$-band data if we stretch the time-axis of the light
curve of SN~1998bw by a ``stretch factor" of $s=0.6$.  The combined
fit of the $V=20.21$~mag host galaxy \citep{cool06} added to the
stretched SN~1998bw $V$-band light curve dimmed by $0.35\;$mag is
shown with the solid line. Our $V$ band light curve of \sn~ peaks at
$\Delta T$ = 10.0 $\pm$ 0.5 days (i.e. 2006 February 28.15 UT and at
$\Delta T$ = 9.7 days in the rest frame of \sn) at an apparent
magnitude $m_V$ = 17.45 $\pm$ 0.05 mag. After correcting for a Galactic
extinction of \av~and host galaxy light contamination, this value
corresponds to a peak absolute magnitude of $M_V$ = $-18.7 \pm$ 0.2 mag
for \sn. The risetime for \sn~is the shortest ever measured for a SN
Ic and is significantly shorter than for \snbw~and \sndh~($\sim 14-16$
days: \citealt{galama98,matheson03}), while \sn~is as almost bright as
SN~1998bw. This is an unusual behavior compared to the sample of
GRB-related SNe (see Fig. 3 in \citealt{stanek05}). We note that there
is a hint of a short plateau phase between $\sim 12-15$ days after the
burst. \it{Thus, we conclude that SN~2006aj is a fast-evolving SN, and
that the SN dominated the light of the OT.}\rm

\begin{figure}
\includegraphics[width=5in,angle=+90]{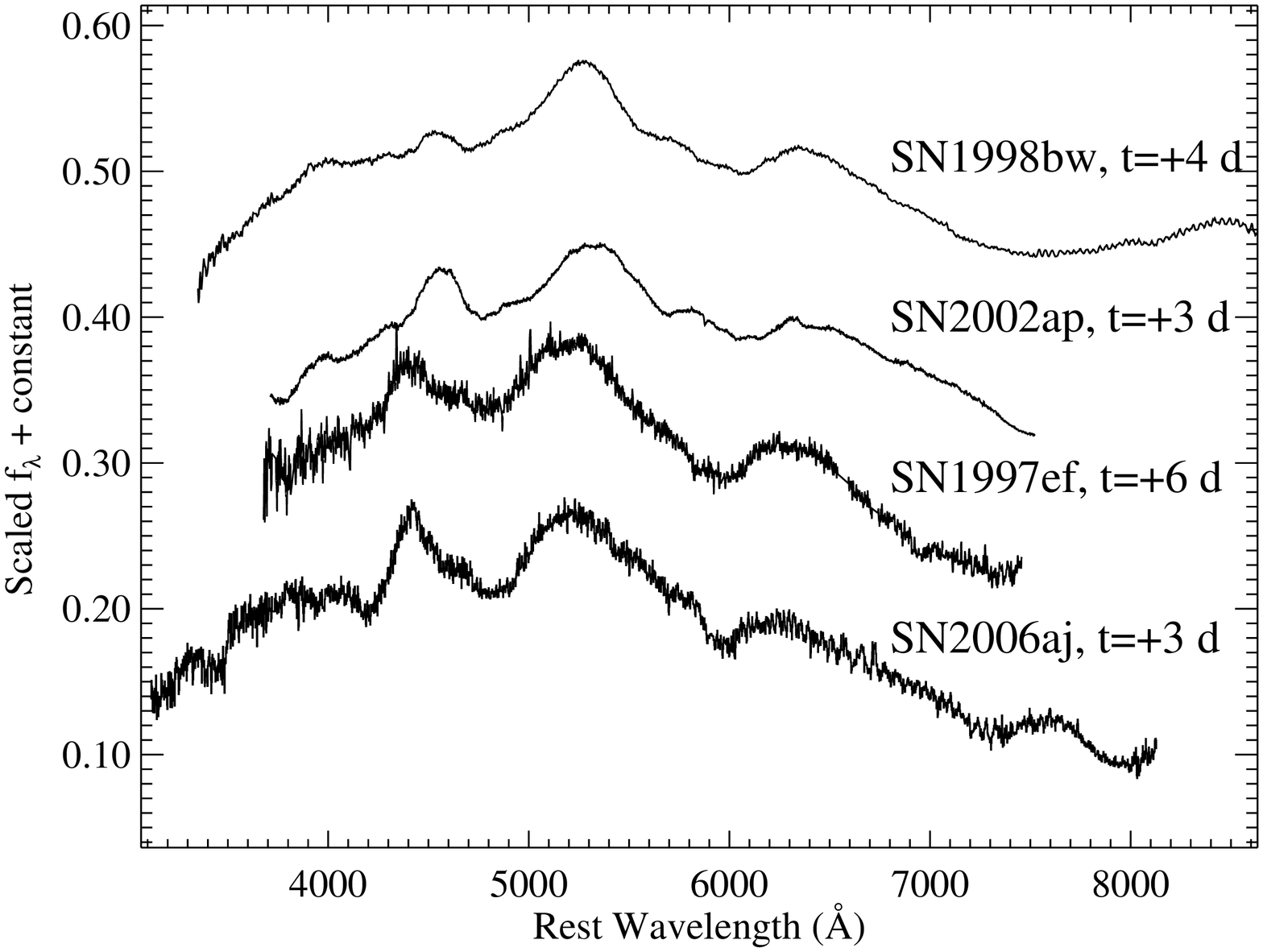}
\caption{MMT Spectrum of \snaj~taken on March 03.00 UT, which
corresponds to $\Delta T =$ 12.85 days after the burst and $\sim$3
days after $V$ maximum.  The broad absorption trough at 5900 \AA~ due
to blueshifted \sitwo~is visible, as well as the broad \ion{Fe}{2}
blends at $\sim$4400 \AA~and \ion{O}{1} $\lambda7774$ at 7300 \AA. For
clarity, the host galaxy emission lines have been removed and the
spectra have been scaled and shifted. From the sample of
representative spectra of broad-lined Type Ic SNe at comparable
phases, it is clear that SN~1997ef and SN~2002ap are better matches
than SN~1998bw. }
\label{mmt_fig}
\end{figure}

In order to study the SN component of \snaj~more closely, we plot in
Figure~\ref{mmt_fig} our MMT spectrum from UT 2006 March 03, at
$\Delta T = $ 12.85 days, thus $\sim$ 3 days after $V$ maximum, when
the SN fully dominates the total light output. The broad absorption
trough at 5900 \AA~(rest wavelength) due to blueshifted \sitwo~is
visible, as well as the broad \ion{Fe}{2} blends at $\sim$4400 \AA,
while no lines of hydrogen or helium are detected. For comparison, we
show spectra of other Type Ic SNe at similar phases: the classical
SN~1998bw \citep{patat01}, the SNe~1997ef (\citealt{iwamoto00}) and
2002ap (\citealt{foley03}, M. Modjaz et al, in preparation). The
spectrum of \sn~exhibits features that are more well-defined and
narrow than those of SN~1998bw, which indicates that the expansion
velocities in \sn~were lower than in SN~1998bw and by extension, in
SN~2003dh. Our earlier spectra of \sn~in addition to this spectrum are
very similar to those of SN~1997ef and SN~2002ap, which are
broad-lined SNe that had no obvious association with GRBs, and are
thought to be due to less energetic explosions than SN~1998bw
\citep{iwamoto00,foley03}. This visual match is supported by
cross-correlating the spectra with our comprehensive library of SN Ic
and GRB/SN spectra (M. Modjaz et al., in preparation) via our
SN-identification algorithm (\citealt{matheson05}, S. Blondin et al.,
in preparation). Thus, the spectra of SNe associated with GRBs seem to
display a certain variety of expansion velocities. Considering both
the light curve and spectral properties of \sn, we conclude that its
expansion velocities lie between that of SN~1998bw and of SN~2002ap,
while a large synthesized $^{56}$Ni mass is needed to explain the
large luminosity in addition to a geometry and ejecta mass that
support the fast escape of photons. We encourage polarization studies
and nebular line spectroscopy to constrain the geometry of the
explosion. Also, late-time observations should give a cleaner window
into the core of the ejecta and help constrain density distribution
and the abundance of nucleosynthesis products.

\subsection{Host-galaxy}

For the adopted luminosity distance of 143 Mpc, the pre-burst
broadband SDSS photometry yields an absolute magnitude for the host
galaxy of \mvhost. This value is less than that of the Small
Magellanic Cloud, a dwarf galaxy with $M_V = -16.9$ mag. We generated
a spectrum of the host galaxy emission lines by averaging the MMT
spectra and subtracting a lower-order fit to the continuum. The
emission line fluxes were measured with the $splot$ task in IRAF and
are given here corrected for Galactic extinction and normalized to
H$\beta$ (where F(H$\beta$)= 0.9 $\pm 0.1 \times 10^{-15}$ ergs
cm$^{-2}$ s$^{-1}$): H$\beta$=1.0, H$\alpha$=3.0, H$\gamma$=0.3,
[\ion{O}{3}]~$\lambda4959$ = 1.2, [\ion{O}{3}]~$\lambda5007$ = 4.0,
\otwo=1.6, and \ntwo$<$0.2 (not detected, 1 $\sigma$ upper
limit). Comparison of the line fluxes with the broadband SDSS
photometry \citep{cool06} indicates that the host galaxy contribution
to the continuum flux is negligible and becomes important only at very
short wavelengths ($3000-4000$ \AA). The relative strengths of the
Balmer lines indicate little host galaxy extinction with $E(B-V) \sim
0.05 - 0.11$ mag. We derive an integrated H$\alpha$ luminosity of
$L(\rm{H}\alpha) = 7.3 \times 10^{39}$ \ergs, which translates to a
current star formation rate of $SFR(\rm{H}\alpha)= 0.06$ \smy
\citep{kennicutt98}. This lower limit is relatively high for such an
underluminous galaxy. In order to derive the metallicity of the host
system, we used the $R_{23}$ iterative diagnostic that involves the
emission line ratios of \otwo,\othree,\ntwo~and H$\beta$
\citep{kewley02}. We derive an ionization parameter of $q \sim 8 \times
10^7$ \cms~and an oxygen abundance of $\rm{log(O/H) + 12} = 8.0 \pm
0.1 $, which corresponds to about $0.15~Z_{\odot}$, assuming a solar
abundance of $\rm{log(O/H) + 12} = 8.86$ \citep{delahaye06}. We note
that the $R_{23}$ diagnostic also possesses an upper branch that
predicts an oxygen abundance of $\rm{log(O/H) + 12} = 8.7 \pm
0.1$. However, this upper branch is excluded by considering the upper
limit on \ntwo~(see Fig. 7 in \citealt{kewley02}). Furthermore,
the higher oxygen abundance would predict a much brighter host galaxy
according to the luminosity-metallicity relationship (e.g.,
\citealt{lee03,tremonti04}).  Thus, we conclude that the host-galaxy
of \snaj~is a low-metallicity, low-luminosity galaxy, very similar to
those of other GRB/SN (e.g., \citealt{matheson03,sollerman05}) and
those of more distant GRBs (e.g., \citealt{fruchter99,lefloch03}).

\section{Conclusions}
We have presented early photometric and spectroscopic data on
\snaj~that clearly establish SN~2006aj as a fast-evolving broad-lined
Type Ic SN with an extremely short rise-time and a large optical peak
luminosity at $z$ = 0.0335. The spectra indicate large expansion
velocities that are smaller than those found in the proto-typical
GRB-related SN~1998bw. The host-galaxy appears to be a low-metallicity dwarf
galaxy.

\acknowledgements We would like to thank the staffs of the MMT
Observatory and FLWO. In particular we would like to acknowledge
C. Tremonti and G. Williams for obtaining some of the spectra, and the
observers at the FLWO 1.2m for obtaining service-
photometry. M. M. thanks T. Matheson for useful comments on the
manuscript and M. Geller, L. Kewley and A. Friedman for helpful
discussions. We thank the $Swift$ team and the observers who provided
their data and analysis through the GCN. Supernova research at Harvard
University has been supported by a grant from the National Science
Foundation, AST-0205808.

\clearpage

\clearpage

\clearpage

\end{document}